\PassOptionsToPackage{unicode}{hyperref}
\PassOptionsToPackage{hyphens}{url}
\documentclass[a4paper,11pt]{article}
\usepackage{lmodern}
\usepackage{amssymb,amsmath}
\usepackage{authblk}
\usepackage{graphics,graphicx}
\usepackage{ifxetex,ifluatex}
\ifnum 0\ifxetex 1\fi\ifluatex 1\fi=0 
  \usepackage[T1]{fontenc}
  \usepackage[utf8]{inputenc}
  \usepackage{textcomp} 
\else 
  \usepackage{unicode-math}
  \defaultfontfeatures{Scale=MatchLowercase}
  \defaultfontfeatures[\rmfamily]{Ligatures=TeX,Scale=1}
\fi
\IfFileExists{upquote.sty}{\usepackage{upquote}}{}
\IfFileExists{microtype.sty}{
  \usepackage[]{microtype}
  \UseMicrotypeSet[protrusion]{basicmath} 
}{}
\makeatletter
\@ifundefined{KOMAClassName}{
  \IfFileExists{parskip.sty}{%
    \usepackage{parskip}
  }{
    \setlength{\parindent}{0pt}
    \setlength{\parskip}{6pt plus 2pt minus 1pt}}
}{
  \KOMAoptions{parskip=half}}
\makeatother
\usepackage{xcolor,hyperref}
\definecolor{darkblue}{rgb}{0.0,0.0,0.5}
\definecolor{darkgrey}{rgb}{0.5,0.1,0.1}
\IfFileExists{xurl.sty}{\usepackage{xurl}}{} 
\IfFileExists{bookmark.sty}{\usepackage{bookmark}}{\usepackage{hyperref}}
\hypersetup{
            colorlinks,breaklinks,
            linkcolor=darkblue,urlcolor=darkblue,
            anchorcolor=darkblue,citecolor=darkblue,
            hidelinks,
            pdfcreator={LaTeX via pandoc}}
\newcommand{\eps}{\varepsilon}

\title{Why we should use Topological Data Analysis in Ageing:\\ towards
defining the ``Topological shape of ageing''}

\author[1,$\ast$]{Tam{\`a}s F{\"u}l{\"o}p}
\author[2,3]{Mathieu Desroches}
\author[4,5]{Fernando Ant{\^o}nio N{\'o}brega Santos}
\author[6,7,$\ast$]{Serafim Rodrigues}

\affil[1]{Research Center on Aging, Geriatric Division,
Department of Medicine, Faculty of Medicine and Health Sciences,
University of Sherbrooke, Sherbrooke, Quebec, Canada}
\affil[2]{MathNeuro Team, Inria Sophia Antipolis Méditerranée, France}
\affil[3]{Université Côte d'Azur, Nice, France}
\affil[4]{Universidade Federal de Pernambuco, Brazil}
\affil[5]{Universitair Medische Centra, Netherland}
\affil[6]{Ikerbasque, The Basque Foundation for Science,
Bilbao, Spain}
\affil[7]{BCAM - The Basque Center for Applied Mathematics,
Bilbao, Spain}
\affil[$\ast$]{Corresponding authors:
\href{mailto:Tamas.Fulop@usherbrooke.ca}{Tamas.Fulop@usherbrooke.ca}, \href{mailto:srodrigues@bcamath.org}{srodrigues@bcamath.org}}

\date{}

\begin{document}
\maketitle

\bigskip

\begin{abstract}
Living systems are subject to the arrow of time; from birth, they
undergo complex transformations (self-organization) in a constant battle
for survival, but inevitably ageing and disease trap them to death. Can
ageing be understood and eventually reversed? What tools can be employed
to further our understanding of ageing? The present article is an
invitation for biologists and clinicians to consider key conceptual
ideas and computational tools (known to mathematicians and physicists),
which potentially may help dissect some of the underlying processes of
ageing and disease. Specifically, we first discuss how to classify and
analyze complex systems, as well as highlight critical theoretical
difficulties that make complex systems hard to study. Subsequently, we
introduce Topological Data Analysis -- a novel Big Data tool -- which may
help in the study of complex systems since it extracts knowledge from
data in a holistic approach via topological considerations. These
conceptual ideas and tools are discussed in a rather informal way to
pave future discussions and collaborations between mathematicians and
biologists studying ageing.
\end{abstract}
\clearpage
\section*{Introduction}
%
\subsection{What is known, what is unknown and what are the unmet needs for
ageing}
Ageing fascinates humans ever since they appeared on earth~\cite{cohen2020a}. 
Why is this the case? Humans are probably the only living
beings that are conscious of their own decay, and they realize that the
end of ageing is death as a natural continuum of life. An almost
invariant hallmark of this decay is that death is preceded by an
accumulation of age-related diseases~\cite{kennedy2014}. If humans
were immortal, then ageing would not bear the same mystery and
fascination. Despite significant advances in science, we still don't
know what is ageing and consequently why humans (and other living
systems) age~\cite{fulop2019}. This lack of understanding is
reflected in the ageing literature, in the sense that the foundational
knowledge on ageing is yet sketchy and even the definition of ageing has
not found a consensus. In part, this is because scientists define ageing
based upon their field of specialization, which also depends on their
capacity to conceptualize and synthesize the complexity of empirical
observations~\cite{cohen2020a}. This lack of consensus questions
whether ageing (as a proper scientific field of study) \emph{per se}
exists. Moreover, it raises doubts whether ageing questions can be
well-formulated and if such questions are amenable to scientific
methodologies, or if ultimately whether ageing research is just a
necessity for scientists to give a sense to their specific niche of
research~\cite{cohen2020b}? Indeed, presently several hundred
theories exist, all of which attempt to define ageing.{~ }However, at
the current stage, it is unlikely that none of these will explain or
ultimately conceptualize ageing~\cite{rose2012,lipsky2015,cannon2015,da2016}.
Moreover, there are so many
aspects of ageing that none of the existing theories can integrate (and
synthesize) them by including molecular, biological, physiological,
functional, social, or psychological aspects. On the other hand, these
uncertainties have led to the generation of data across various scales,
starting from molecular through the cellular to the organismal level.
These data exist most of the time in silo and serve mainly to explain
correlation but not to understand the holistic causal relationship of
the data defining ageing. However, some attempts exist to unify the
ageing concept and try to find a causal relationship among the different
data conglomerates or pathways. We subsequently provide three examples,
which are the most advanced ones to conceptualize ageing.{~}

The first example is the \emph{epigenetic clock,} described by Horvath~\cite{horvath2013}.
This clock predicts biological age and provides a
collection of innate biological mechanisms that give rise to age-related
DNA methylation changes that underlie highly accurate DNAm age
estimators, and thus provides a tentative answer as to why we age~\cite{horvath2018}.
This clock was derived via traditional
statistical methods such as the \emph{metaAnalysis} R function in the
WGCNA R package to measure pure age effects and ANOVA for analysis of
variance for measuring tissue variation~\cite{horvath2013}. Since then,
several epigenetic clocks have been described~\cite{belsky2018} with
the aim to determine the biological age compared to the chronological
age (Hannum, PhenoAge and GrimAge)~\cite{li2020,moskalev2020}.
The second example is the \emph{dysregulation} concept of Cohen et al.~\cite{milot2014,
cohen2015}, which employs a more advanced
computational methods (e.g. statistical distance, D\textsubscript{M},
Principal Component Analysis) to explain the complexity of ageing. There
is now clear evidence that physiological dysregulation-\/-the gradual
breakdown in the capacity of complex regulatory networks to maintain
homeostasis-\/-is an emergent property of these regulatory networks, and
that it plays an essential role in ageing. It can be measured merely
using small numbers of biomarkers~\cite{cohen2016,mitniski2017,chmielewski2020}.
The third and most recent example can be called the
\emph{multi-level process} of ageing, which is being led by the research
group of Franceschi~\cite{whitwell2020}. In this approach biological
ageing is considered as a complex process involving multiple biological
processes, which can be understood theoretically by considering them as
an individual network (e.g. epigenetic networks, cell-cell networks, and
population genetics). Mathematical modelling (e.g. via multi-layer
networks~\cite{kivela2014}) allows the combination of such networks
so that they may be studied as a whole, to better understand how the
so-called ``seven pillars of ageing'' combine~\cite{kennedy2014}. This
approach also has the potential of generating hypotheses for treating
ageing as a condition at relatively early biological ages~\cite{whitwell2020}.
These examples highlight the unmet need for more advanced
computational tools to further dissect the complexity of ageing from the
multitude of data sets gathered during these last decades~\cite{zhavoronkov2019}.
Ideally, these tools should treat ageing data as a
problem (and in a holistic sense), extract causalities, explain the data
mechanistically and ultimately define targets for ageing interventions~\cite{zhavoronkov2019}.

Several recent attempts have been put forward recently to determine the
different components of the biology of ageing recognizing that indeed
aging is a multicomponent complex process. For instance, \emph{the nine
hallmarks of ageing} by the group of Kroemer~\cite{lopez2013}
has been proposed. These hallmarks are conceptualized by noting that
ageing is led by molecular and physiological deregulation and the
subsequent appearance of pathological consequences such as diabetes,
cancer, cardiovascular diseases. Specifically, the hallmarks are genomic
instability, telomere attrition, epigenetic alterations, loss of
proteostasis, deregulated nutrient sensing, mitochondrial dysfunction,
cellular senescence, stem cell exhaustion, and altered intercellular
communication~\cite{lopez2013}. Moreover, these studies have
tried to define some of the components of these hallmarks but
unfortunately, most of these are at a phenomenological level. Thus, this
remains very descriptive and does not preclude any causality, and even
the interaction and interconnection between these hallmarks are
conjunctural only. Moreover, no hierarchy has been established
adequately among these hallmarks in the sense to provide causal
interactions to generate a holistic/global/integrated response of the
entire organism during aging. The ultimate aim was indeed to find
targets for intervention to cure/modulate/influence or reverse ageing to
avoid or at least decrease the emergence of the above-mentioned
age-related diseases. In this sense, ageing is considered a disease that
requires a cure primarily to decrease the burden of age-related diseases~\cite{bulterijs2015},
however considering all biological aspects of
aging it appears that ageing cannot be considered as a disease~\cite{fulop2019}.{~}

This view has been adopted and refurbished by the so-called geroscience,
which clearly conceptualizes ageing as a pathological phenomenon that
should be controlled to decrease age-related chronic diseases~\cite{kennedy2014}.
Moreover, geroscience conceptualizes ageing by the
interaction of the seven pillars such as metabolism, epigenetics,
inflammation, adaptation to stress, proteostasis, stem cells and
regeneration, macromolecular damage. These are somehow more broad
processes than that defined by the nine hallmarks, but most probably
there are many overlapping between the two classifications. At least the
aim is the same, which is to use ageing as a target to decrease
age-associated diseases and prolong health span. While this is an
important goal, the hallmarks/pillars classification will only enable
the derivation of methods that intervene separately on individual
hallmarks since most interactions are pair-wise (and not genuinely
causal) and thus holistic interventions on ageing (and associated
emergent processes) are likely to fail.{~}

Furthermore, it is crucial to stress that many of these
hallmarks/pillars are mainly derived from animal models of ageing.
However, this is now quite accepted that none of these models alone may
not represent entirely human ageing either because of the longevity or
the complexity or the environmental stresses. As such, it is possible
that the described hallmarks/pillars are simply the surrogate of the
specific animal models/pathways even if they are presented as universal
underlying biological phenomena across the whole animal kingdom.{~}

Yet another problem is that the hallmarks/pillars are only rarely
related to the dynamic physiological or functional aspects of ageing~\cite{belsky2018}
even if there are some attempts to connect to
physiology in view of better determining the mortality~\cite{li2020}.
Thus, it is almost impossible to state what is the potential
impact of these models on the ageing of humans, and in the best-case
scenario, these may only serve as surrogate data. Even longitudinal
studies have so far failed to answer these crucial concerns. As a
biological process and as much as it can exist if there is no
physiological or functional repercussion, ageing \emph{per se} is just
an adaptation to the passage of time, as was suggested e.g. for
immunological changes~\cite{franceschi2017,fulop2018}. For
example, grey hair may represent an ``ageing'' phenomenon, but it has no
real consequences except that probably cell senescence is the underlying
process. In contrast, the decrease of the hepatic blood flow which
occurs with ageing, probably ultimately due to mitochondrial
dysregulation (metabolic, oxidative, inflammatory), is not measured and
rarely considered, yet it may have serious consequences on
detoxification. Thus, ageing is one of the most intriguing processes
that humans ought to understand through life since what can be
phenotypically observed as ageing is not necessarily associated with the
real complex mechanisms of ageing. In other words, we can observe
chronological age, but we cannot yet dissect the underlying biological
processes of ageing that occur along time. Human data related to some
dynamic functional processes should, therefore, be acquired in order to
unravel the importance of these processes for ageing. Furthermore, one
could determine in this way the biological age of individuals. Indeed,
the above-mentioned epigenetic clock tried to fill this gap~\cite{horvath2013,horvath2018}.

Another widespread approach to ageing is its consideration as only a
deleterious process given that it is inevitably resulting in death.
However, many phenomena that are described, measured, or considered as
related to ageing may be simply adaptative processes designed to
optimize the functioning of the organism at a specific moment and
circumstance of life related to some evolutionary necessity~\cite{franceschi2017,fulop2018}.
Until now, all described hallmarks
are considered to some extent as causes of ageing however they cannot be
considered as \emph{a priori} deleterious and being targeted for
intervention to ``rejuvenate'' the system since their effect are not
determined from the viewpoint of species or individuals' evolution. In
the immune system, the best example is the activation state of the
innate immune system with ageing which may be considered as the
\emph{inflamm-aging}~\cite{franceschi2017} but, in contrast, may also
be considered as an evolutionary defence mechanism against repetitive
challenges from both outside and inside in order to support a better
readiness of the organism~\cite{fulop2016}. As long as we cannot
fully understand this phenomenon in the complexity of the immune
response changes related to ageing, we cannot really decide what the
role of this resting overactivation is, and how we should intervene if
any intervention is necessary.

Thus, until now many data have been gathered at various levels, but no
efficient tools exist that are able to conceptualize in a causal manner
all these processes, either molecular or cellular, tightly linked to
physiological or clinical data. Interdisciplinary Researchers and
clinicians should closely collaborate to address and hopefully answer
the following key questions: \emph{``What is the meaning of these basic
hallmarks of ageing on the physiological/clinical performance of the
older subjects? And would targeting them change something in the
trajectory of an older individual subject for a better healthspan?''}
Addressing these key challenges will be the foundation of precision
medicine for older subjects. A thorough understanding of ageing thus
implies an integrative, complex systems framework where lower-level
mechanisms can have direct impacts (e.g. mutations causing cancer or
cellular senescence), or indirect impacts via higher-level processes
(e.g. impacts of inflammation on atherosclerosis). Higher levels of
organization

{~}can also have their own independent mechanisms (e.g. tooth wear).
This framework would help explain the diversity of ageing patterns
across the tree of life, with both ``public'' and ``private''
mechanisms. We should acknowledge that numerous attempts exist to use
artificial intelligence (AI) tools with the aim to better understand and
conceptualize the complexity and emergent nature of ageing. Yet it seems
that they still catch only the fragments of the causes of the whole
dynamic ageing process. Thus, developing and using new AI tools are
crucially needed to deeply understand the causal relationship of the
many data gathered on ageing, with the overarching goal to be able to
understand the quintessence of ageing and consequently modulate it in a
holistic manner.{{~}}

\hfill\break

\subsection*{Can ageing be understood?}
Living cells (about 30 trillion in humans) are complex systems that
encompass anatomical structures, information maintained by DNA,
bioenergetics and information processing mediated by multi-omic
subcellular and cellular machinery. A disruption of the underlying
cellular mechanisms contributes to ageing and diseases. A timely and
significant question is whether it is possible to fully explain or
dissect the biological processes of ageing from multi-omics data and
other relevant biological markers? An affirmative answer to this
question has far-reaching medical implications and the tantalizing
prospect of partially reversing ageing and diseases that result from
ageing. However, this question rapidly encounters non-trivial scientific
hurdles, since biological structures and processes of living organisms
are emergent and strongly coupled across different spatio-temporal
scales (i.e. from microscale to macroscale)~\cite{dumont2014}.
By emergent, we mean that a macroscopic property (structure or
process) is not merely the sum of its microscopic property (e.g. water
can be in the liquid state and can wet a cloth, but a single water
molecule has no such macroscopic property). In the case of ageing, one
is interested in explaining the macroscopic phenotypes in terms of the
measured microscopic data (e.g. omics). Moreover, by strong coupling
between scales, we mean that there are feedback coupling loops between
scales (mostly unknown) in such a way that emergent properties
(multiscale information) induce changes onto the microscopic properties
and vice versa. As a consequence, dissecting biological processes, in
particular, those associated with ageing, into building blocks and
studying these building blocks separately (i.e. via the traditional
scientific reductionist method) is bound to fail. Novel scientific
methods, particularly those that have a holistic (in the sense of
systems biology) approach, are necessary to make sense of complex
systems. These ideas, and in particular the extent to which one can
dissect a complex system, are best understood in the context of a
classification process, developed within the field of Theoretical
physics. A given system can be of one of the four types: \emph{Strong
reduction}, \emph{Supervenience}, \emph{Contextual emergence},
\emph{Radical emergence} (see Fig. 1)~\cite{bishop2006,bishop2020,butterfield2011}.

\hfill\break

\begin{figure}[!t]
\centering
\includegraphics[width=0.7\textwidth]{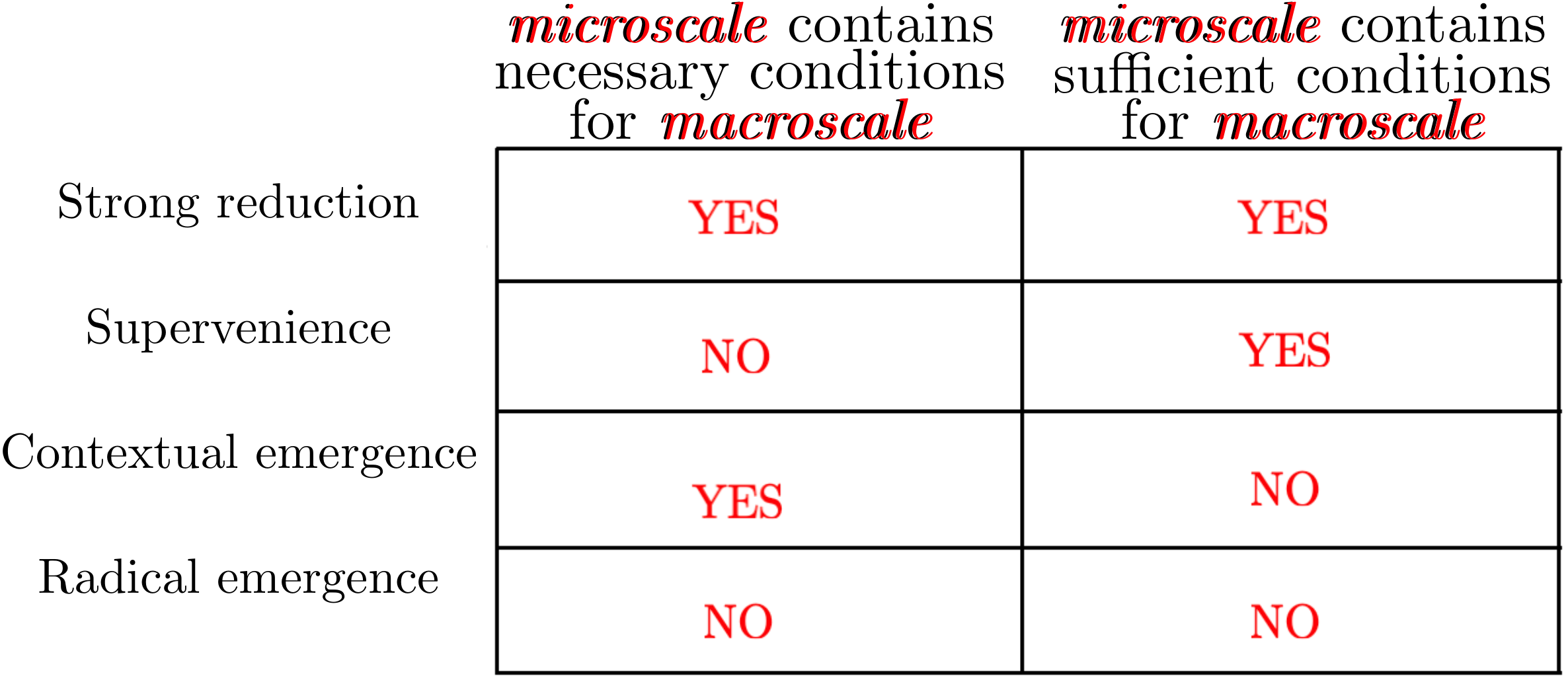}
\caption{\emph{Classification map of physical and biological systems
from theoretical physics and philosophy of science point of view. There
are mainly four types of systems, which are classified based on the
amount of information available at the microscale to explain the
macroscopic observations. For the case of ageing in biological systems,
the class is unknown, but the aim is to describe the macroscopic
observations of ageing in terms of the microscopic biological
mechanisms.}}
\end{figure}

\emph{}~\\

The simplest class is the so-called \emph{strong reduction,} in which a
system can be fully dissected into building blocks. This is only
possible if at the microscopic scale the system contains necessary and
sufficient conditions (these rigorously expressed mathematically) to
explain the macroscopic properties of the system. Ultimately, these
conditions permit to dissect the system and explain it in terms of its
elementary building blocks (i.e. the system is the sum of its parts).
Examples within this class are thermodynamic closed systems, such as a
container containing gas, where the external application of temperature
(or manipulation of any thermodynamic parameters) leads to changes of
the macroscopic state of the system to either gas, liquid or solid (see
Fig.{~}2, Panel A).{~ }These macroscopic states (and associated dynamics)
can be fully understood (mathematically) in terms of the microscopic
atomic activity. This understanding is possible due to the existence of
necessary and sufficient physical and mathematical conditions
(discovered in the field of Mathematical-Physics), namely: Boltzmann's
\emph{Propagation of molecular chaos}, \emph{zeroth law of
thermodynamics} and so-called \emph{Kubo-Martin Schwinger (KMS) states}
~\cite{ehrenfest1959,gottlieb2003,haag1967,kubo1957,martin1959,villani2002}{. }We shall not
delve into the technical details of these conditions, but the intuitive
idea is that the interactions between the gas particles are elastic and
dissipate (memoryless) and so the effect of each particle on the overall
gas is negligible. As a consequence, it is possible to coarse grain and
average (i.e. via a statistical distribution) the activity across
various scales. This idea leads to the derivation of a \emph{macroscopic
time-evolution equation} and \emph{thermodynamic observables} (i.e.
functions that depend on thermodynamic variables: \emph{pressure},
\emph{volume}, \emph{temperature}, \emph{Avogadro constant}), which
predict respectively, the time dynamics of the system and state phase
transitions or tipping points (e.g. changes for solid to liquid).
\begin{figure}[!t]
\centering
\includegraphics[width=0.7\textwidth]{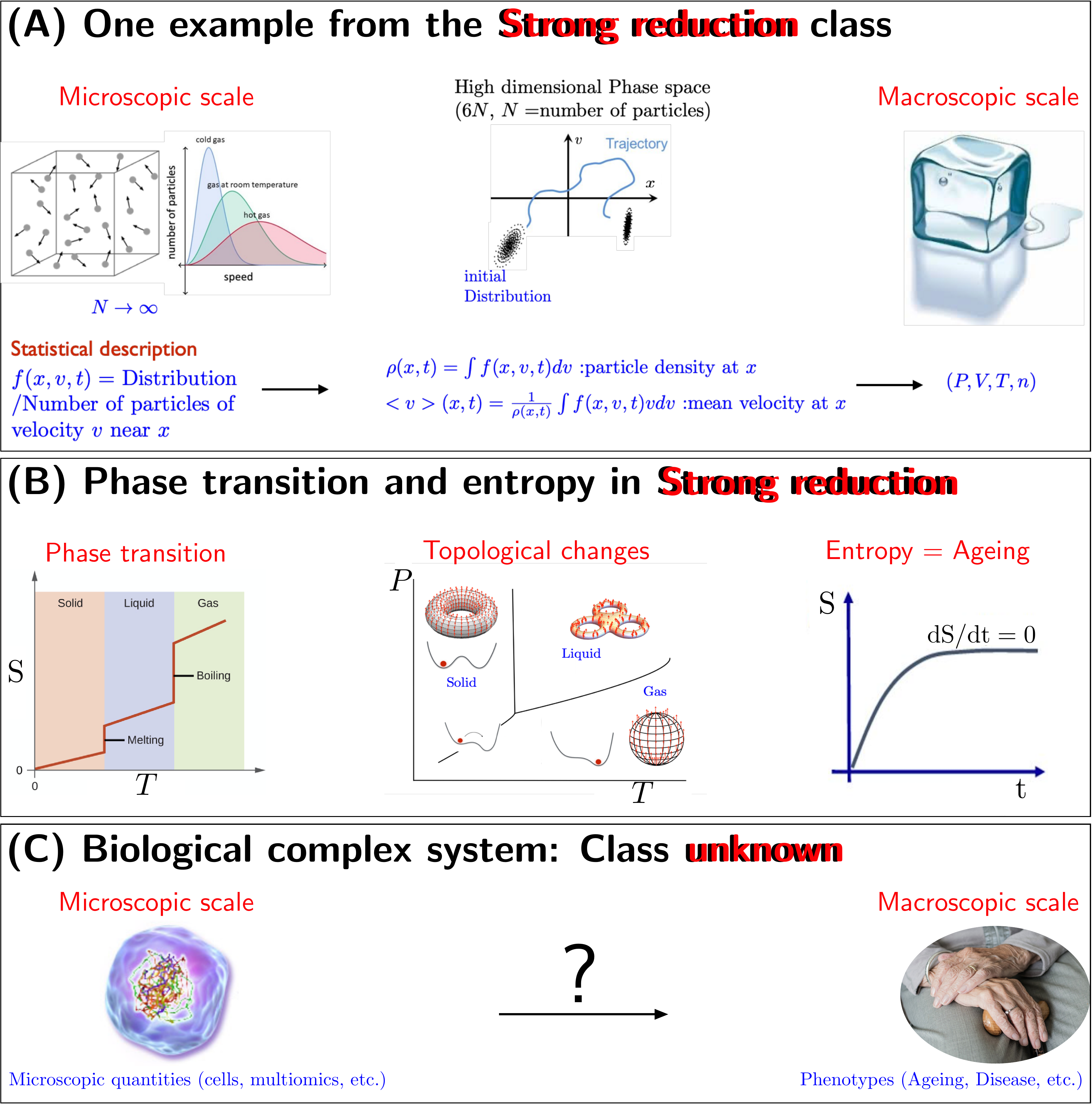}
\caption{\emph{Strong reduction systems vs complex systems. A: depicts
an example of a strong reduction system. The traditional point of view
of the scientific approach is that systems can be broken up into micro
building blocks and then the assembling of it provides an explanation of
the entire system. This approach has been successful in explaining many
physical systems, such as the depicted example of a closed thermodynamic
system of a gas contained within a container. The macroscopic states of
matter (solid, liquid, gas) can be predicted by dynamic equations and
functions, which explain how the building blocks interact. B. Phase
transitions of states of matter can also be predicted by rapid changes
observed from thermodynamic functions (singularities), transitions in
the energy functions in undergoing phase transitions. This processed is
induced by necessary topological changes of the configuration of
particles in the space of position and velocities. The ageing of these
closed systems is simply described by Entropy. C: In contrast, the
biological systems are complex, where some emergent features (or
phenotypes) may not just be a sum of the building blocks. This occurs
when there are strong feedback and coupling across spatial-temporal
scales of the systems. Therefore, the class of biological systems is yet
unknown.}}
\end{figure}
Noteworthy, the macroscopic equation encodes a statistical distribution
(specifically Maxwell-Boltzmann distribution) of atomic particles and
the ultimate reason as to why it is possible to explain the macroscopic
properties from the microscopic ones is because there is a one-to-one
mapping between the five parameters of the distribution and five known
conservation laws of physics. Changes between states (e.g. solid, liquid
and gas) occur because the energy landscape and microscopic cloud of
atoms radically reorganize (i.e. the overall topology or configuration
of atoms in the space of possible positions and velocities change, see
Fig 2. Panel B) as the thermodynamic parameter(s) vary. Under these
radical changes, \emph{thermodynamic observables}, exhibit a singularity
(sharp changes; see Fig.{~}2, Panel B). Moreover, the ageing process of
such closed systems is fully described by the second law of
Thermodynamics, which states that the entropy of the system always
increases (see Fig.{~}2, Panel B). The second class, \emph{supervenience},
corresponds to systems whereby the microscopic scale only contains
sufficient conditions to explain macroscopic properties. The third
class, \emph{contextual emergence}, corresponds to systems in which the
microscopic scale contains only necessary conditions to explain
macroscopic properties. An interesting example of contextual emergence
lies in the relation between topology and phase transitions in
theoretical physics. In fact, under certain conditions, a topological
change in the configuration space of a Hamiltonian system emerges as
necessary conditions for the occurrence of phase transitions, but not
sufficient~\cite{franzosi2004}.{~ }The last class,
\emph{radical emergence}, are systems without sufficient and necessary
conditions at the microscopic scale, and thus the macroscopic properties
are purely emergent and impossible to explain. 
Presently, there is no consensus to which class life (biological systems) belongs to; 
see Fig.{~}2, Panel B. These are open systems (e.g. cells interact with the
external world and extract energy via the ATP system), thus conservation
of laws of physics do not apply (at least at the microscale). Moreover,
there are strong coupling across different spatio-temporal scales, and
some processes seem emergent, then it makes it hard to study. However,
the scientific hope is that these complex systems are mixed, either
\emph{supervenient} and/or \emph{contextual emergent,} which would
provide a basis to study life and ageing with some level of mathematical
and scientific rigour. A fundamental question is whether we can develop
scientific tools that could shed light onto complex systems, even if it
is only partially? The subsequent section discusses a novel Big Data
tool, the so-called \emph{Topological Data Analysis}~\cite{carlsson2009,rabadan2019,wasserman2018},
which has the potential to
substantially advance our understanding of complex systems~\cite{petri2013},
namely, in the context of the present manuscript, to ageing
and diseases.

\section*{Topological Data Analysis: A novel tool to study complex
systems}
Topological Data Analysis (TDA) is a relatively novel and revolutionary
Big Data method, with origins in pure mathematics~\cite{hatcher2002} and
which has been recently further developed in applied fields, e.g.
physics and computer science~\cite{carlsson2009,wasserman2018}. In
contrast to existing AI (Artificial Intelligence)~\cite{lecun2015}
and Big Data tools~\cite{sagiroglu2013} that determine numerical
summaries/patterns (and typically incomplete), TDA extracts knowledge
from data in the form of shapes (technically \emph{topological
invariants)} and non-trivial relations between shapes. Specifically, it
extracts from noisy multiscale high-dimensional data the intrinsic
global structural shape (holistic descriptor) of a complex system~\cite{blevins2020}.
Since TDA has a rigorous theoretical
foundation, it is ideally suited for extracting complex causal
interactions across spatio-temporal scales and consequently has the
potential to establish non-trivial relations between microscale and
macroscopic emergent processes. Thus, TDA has the potential of extending
our knowledge of living systems and, in the context of the present
manuscript, dissect (to some degree) the processes underlying ageing and
associated diseases. The effectiveness of TDA has been shown in several
applications, including biological and clinical data~\cite{sasaki2020,rabadan2019}.
To cite a few examples: 1. TDA applied to a database
featuring multi-omics data of type-2 diabetes patients, discovered a
third unknown form of diabetes~\cite{li2015}; 2. Applied to malaria
medical records featuring transcriptome data and phenotypes like
temperature, parasite density, etc., it tracked disease progression and
characterized resilient patients~\cite{torres2016}; 3. Applied to
clinical data of Asthma patients featuring physiological and
inflammatory parameters it discovered key endotypes~\cite{hinks2015,hinks2016}.
As a consequence, {it} is envisaged that TDA will
become a fundamental method for precision medicine.

\begin{figure}[!t]
\centering
\includegraphics[width=0.7\textwidth]{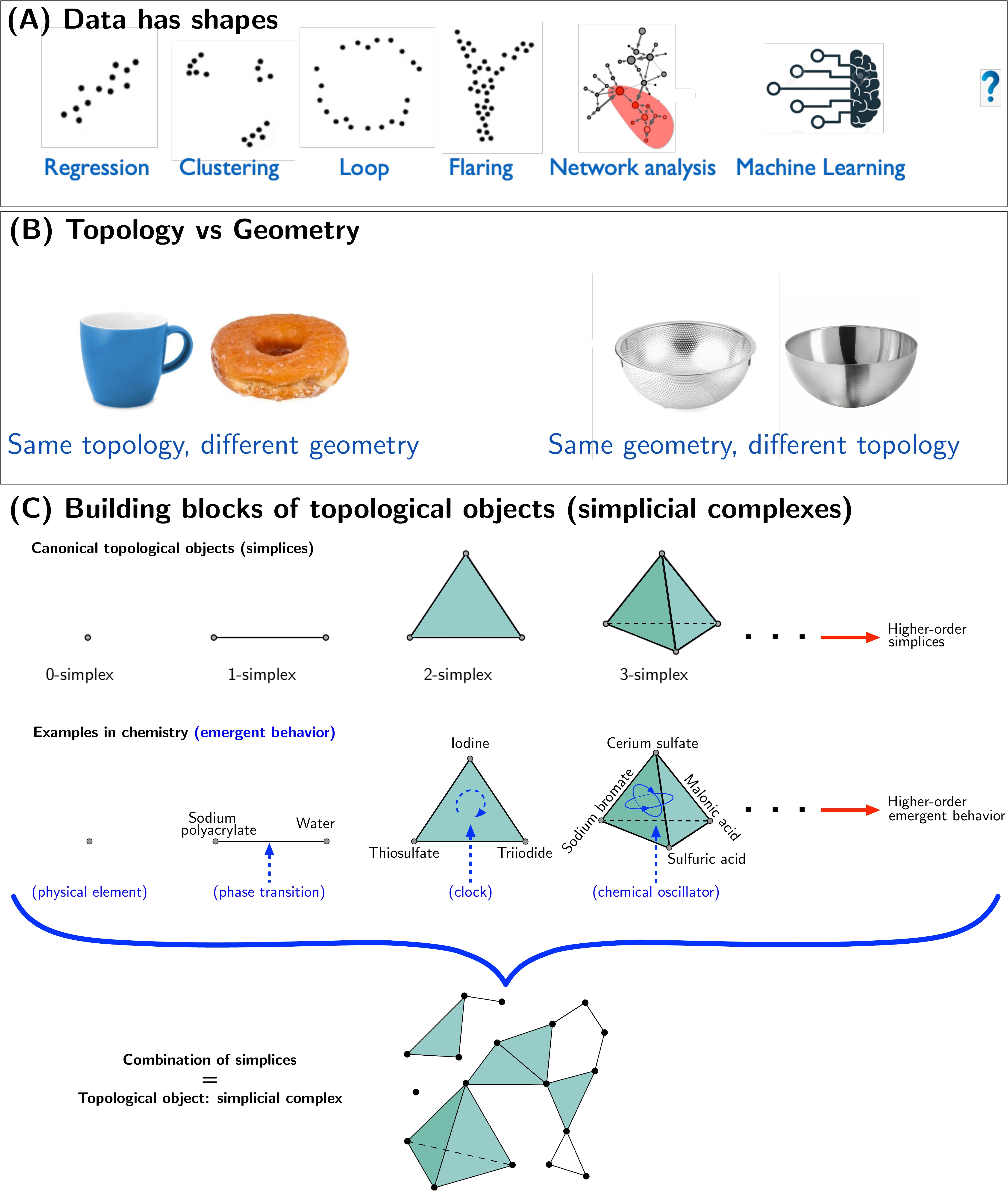}
\caption{\emph{A: In general data has an intrinsic shape (pattern) that
encodes useable information. Different methods have been developed but
somewhat limited. Thus, the question is if there is a more general way
to mine useable information. B: There is a difference between Geometry
and Topology. Geometry is concerned with lengths and angles, while
topology is concerned with shapes that remain invariant under elastic
deformation. C: Topological objects (simplicial complexes) are
constructive and built from building blocks called simplices. Simplices
in general represent higher order relations (e.g. emergent properties).
For example, 2-simplex is the face a triangle, which explains the
emergence of a chemical clock after mixing three chemical elements (i.e.
data points).}}
\end{figure}

To make sense of TDA, it is first important to note that measured data
has an intrinsic shape (technically, it has a \emph{topology}). This
concept is exemplified in Fig.{~}3 (Panel A), where scattered data (i.e.
\emph{point cloud}, a distributed discrete collection of data points)
encodes a pattern from which we may wish to decode and synthesize
mathematically. However, there are infinitely many ways in which data
can be distributed, and this raises the question of whether there exists
a general algorithm for extracting these mathematical patterns?
Traditional data analytical approaches implement specific algorithms to
extract specific patterns within data, which are sub-optimal. To
illustrate this point, consider the following two examples: 1. If the
data is scattered along a line, then a regression method that fits a
line (i.e. a functional relationship between variables) is applied. The
approach (algebraic modelling) is only suited for systems with low
degrees of freedom (i.e. small number of variables, parameters and
simple relationships). Moreover, algebraic fitting requires a high
degree of approximations. This means the degree of understanding that
the scientist extracts from the model (explanation of the mechanism
behind the data) degrades for complex data. Moreover, this approach
always requires an initial hypothesis (a kind of supervised analysis).
2. If the data is scattered into clusters, then a suitable clustering
algorithm can be used. Unfortunately, there are many clustering
algorithms due to the necessity of optimizing an objective function of
some kind and, as a consequence, none of these works so well since there
are no known optimal objective function. Beyond these two examples data
can be distributed in more arbitrary ways (e.g. circle, flaring or more
complicated ways), thus one quickly realizes that deriving a new
algorithm for each data set is not tenable. Recent, Big Data and AI
approaches (e.g. machine learning and network theory) attempt to address
some of these deficiencies (e.g. unsupervised learning), but
unfortunately these solutions are only partial and are unable to
determine causal relations (e.g. machine learning)~\cite{baker2018}.
TDA overcomes these deficiencies in four steps as follows: 1. It first
defines what is meant by shape by equating it to topology rather than
geometry (see Fig.{~}3, Panel B). Topological objects are those that are
invariant (or remain unchanged) under continuous deformations/stresses
(i.e. elastic deformation like a rubber band that is stretched but not
torn apart). For example, consider a coffee mug (see Fig.{~}3, Panel B),
which has one hole in the handle of the mug. Consider also a table
plate, which has no hole. These two objects are topologically different
since it is impossible to continuously deform (elastically) a coffee mug
and transform it into a table plate (and vice versa). However, it is
possible to continuously deform a coffee mug into a torus (e.g. donut)
since both objects have a hole. Thus, these objects are Topologically
the same because they have the same invariant, in particular, the same
number of holes. To make this point clear, see the second example (see
Fig.{~}3, Panel B). Although the two bowls have the same geometric look,
they are topologically different because one of the bowls have numerous
holes. Thus, it is possible to elastically deform one into the other. 2.
Second, TDA notes that a complex system can be associated with a global
structure (global shape or global topological object, technically called
\emph{simplicial complex})~\cite{salnikov2018}. This structure
encodes data that can be decomposed into canonical patterns (canonical
topological objects, technically called \emph{simplices}); see Fig.{~}3, 
Panel C. In analogy, these canonical topological objects can be seen as
the basic constituents of a system, such as atoms within the periodic
table (in chemistry), or the fundamental musical chords from which any
music can be composed. The simplest canonical object is a single data
point (denoted the \emph{0-simplex}). Subsequently, a pair-wise relation
between two data points is denoted a 1-simplex. The study of 1-simplices
is essentially network or graph theory. That is, graph theory is built
upon pair-wise relations only~\cite{battiston2020}! The next level
of complexity, the \emph{2-simplex}, determines the case when three data
points are simultaneously participating in a causal relationship, from
which emerges a new object (represented as the face of a triangle). An
example of such a scenario is the emergence of a chemical clock when
three chemical compounds interact (i.e. Iodine, Thiosulfate and
Triiodide). Clearly, the 2-simplex is already beyond the standard
network and graph theory! Higher-order simplices (n-simplex)~\cite{lambiotte2019,salnikov2018}
follows the same rationale,
which is to represent the emergence of novel higher-order objects (see
Fig.{~}3, Panel C). Finally, an arbitrary \emph{simplicial complex} (i.e. a
global topological object) can be formed by combining suitable simplices
(see Fig.{~}3, Panel C). 
\begin{figure}[!h]
\centering
\includegraphics[width=0.7\textwidth]{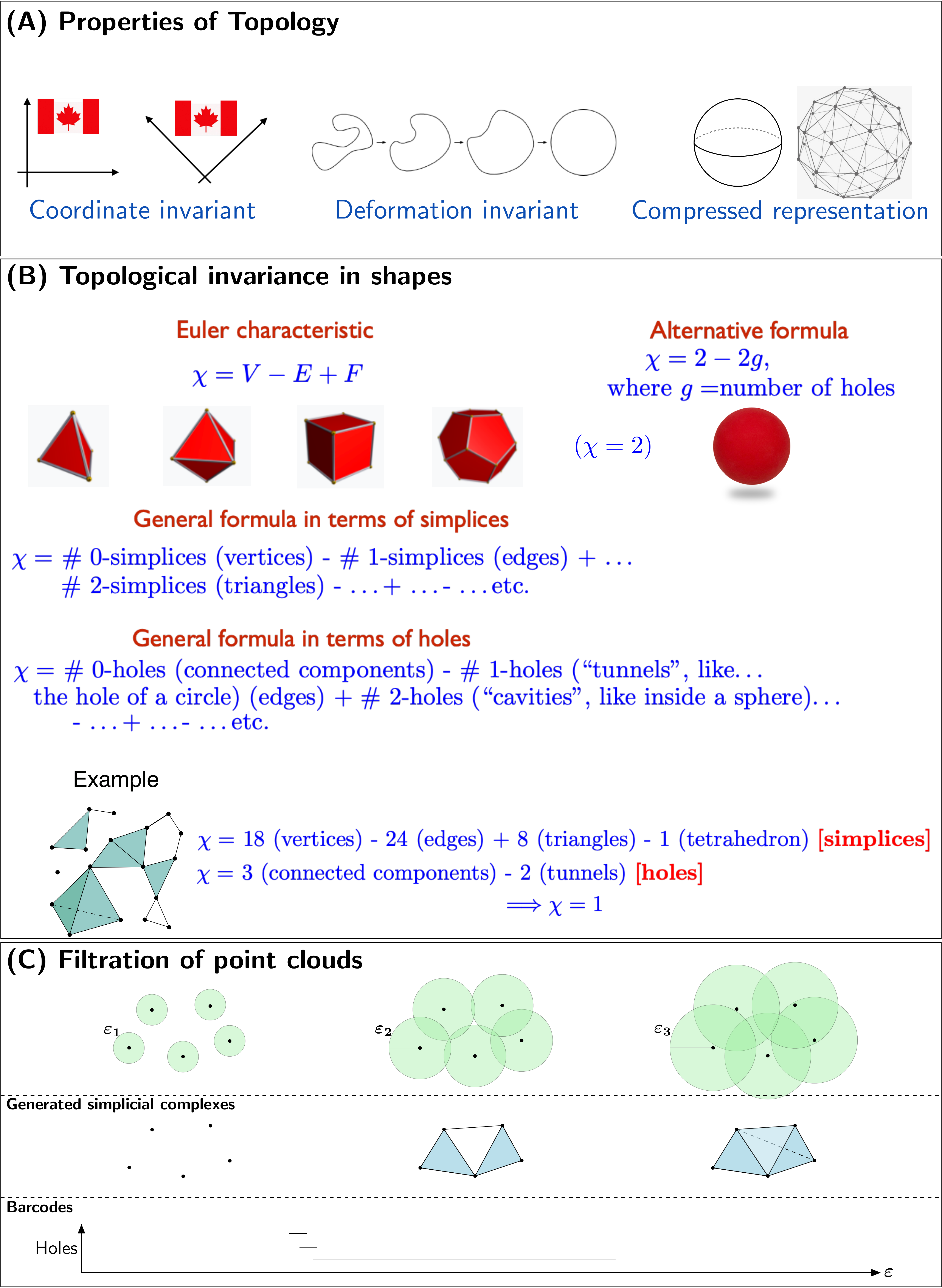}
\caption{\emph{A: Three fundamental properties of Topology. B: Platonic
solids have the same topological invariance as a sphere. Specifically,
for the above examples we have the following. Tetrahedron has 4
vertices, 6 edges and 4 faces, which equates to Euler Characteristic
(EC) = 2. An Octahedron has 6 vertices, 12 edges and 8 faces, hence
EC=2. A cube has 8 vertices, 12 edges and 6 faces, hence EC=2. A
Dodecahedron has 20 vertices, 30 edges and 12 faces, hence EC=2. Equally
the Sphere has EC = 2. The Euler characteristics of a simplicial complex
is also constructive as shown in the example. C: Persistent homology is
one important TDA algorithm that mines topological patterns from data
and the algorithm uses a process called filtration of point clouds. The
filtration consists in positioning balls with centre at the data points
and with a certain radius (defined with respect to some similarity
measure) and continuously expand these balls (i.e. increment the
radius). The intersection of the balls generates simplicial complexes
and the topological invariances can be tracked with barcodes.}}
\end{figure}
3. Crucially, topological objects (i.e. simplices
or simplicial complex) satisfy three fundamental properties, namely,
\emph{Coordinate Invariant, Deformation Invariant,} and \emph{Compressed
representation}~\cite{carlsson2009} (see Fig.{~}4, Panel A) and have an
associated unique signature, technically, \emph{topological invariance}
(see Fig.{~}4, Panel B). \emph{Coordinate invariant} means that the
topological properties of the data are independent of the coordinate
system. This property is in contrast to a large class of Big Data
techniques (e.g. Principal Component Analysis). \emph{Deformation
invariant} is a by-product of the aforementioned definition of a
topological object, which states that they are invariant (or remain
unchanged) under continuous deformations (i.e. elastic deformation like
a rubber band that is stretched but not torn apart). This property makes
TDA robust to noise, in the sense that small perturbations of the data
will not change the underlying topological properties of the object
encoded within the data~\cite{blevins2020}. \emph{Compressed
representation} means that a given topological object can be represented
by a minimal number of data points, as long as the data points do not
change the topology of the object. \emph{A topological invariant} is a
unique signature or measure that quantifies the underlying pattern
(topological object) within the data. The idea of topological invariance
can simply be understood via the example shown in Fig.{~}4, Panel B. As
shown, the four platonic solids (tetrahedron, octahedron, cube,
icosahedron) and a sphere, although possessing different geometries, are
essentially the same topological object (i.e. one can deform elastically
one into another and indeed the four platonic solids are compressed
representations of the sphere). The equality between these objects is
quantified via the \emph{Euler characteristic} (in its simpler form
equates to vertices -- edges + faces; see Fig.{~}4, Panel B), which
provides the topological invariance. To clarify, consider the
Tetrahedron, which has 4 vertices (i.e. the data points), 6 edges
(relation between data points), 4 faces (higher-order relations); the
corresponding Euler characteristic is 2. Equally, applying the Euler
characteristic's formula to the octahedron, the cube or the icosahedron,
also yields in \emph{Euler characteristics} equal to 2. Moreover, the
sphere also has a \emph{Euler characteristic} of 2. This can be
understood by the alternative formula in Fig.{~}4, Panel B, which counts
the number of holes (in the case of a sphere it is zero). Alternatively,
one can imagine a sphere being cut in half through a \emph{great circle}
that goes through the north and south pole (i.e. 2 vertices), which
forms two \emph{meridians} (i.e. 2 edges) and split the surface of the
earth into two half-spheres (i.e. 2 faces). Therefore, the \emph{Euler
characteristic} is indeed 2. This short exercise emphasizes the
importance of identifying topological invariances within data (a
fundamental signature), which gives us confidence about the underlying
pattern within data even under noise perturbations. In the exercised
example, adding or removing data points (i.e. noise perturbations) to
the sphere will only result into either a smoother sphere or a platonic
solid, all of these having the same invariant measure, hence equivalent.
More generally, the \emph{Euler characteristic} is an alternating sum of
simplices or equivalently holes (see Fig.{~}4, Panel B). Consequently, a
given simplicial complex is a composition of its constituent building
blocks (e.g. simplices and holes) quantified by the \emph{Euler
characteristic} (see example in Fig.{~}4, Panel B). Simplicial complexes
are in a way discrete analog of multidimensional surfaces. Since
surfaces have multidimensional holes, simplicial complexes also do~\cite{zomorodian2005}.{ }
4. The fourth and final step develops a
computational machinery to search (or extract) simplicial complexes from
a distributed discrete collection of data points. There are different
methods and approaches in TDA, to name a few, \emph{Persistent
homology}, \emph{Reeb Graph}s, \emph{Mapper algorithm}~\cite{singh2007,geniesse2019}\emph{.}
Herein we will only discuss the
method of \emph{Persistent homology} since it is general enough to have
an idea of TDA and has successfully been applied to biological data. The
first step in persistent homology is to consider a similarity measure
between data points (e.g. correlation, causality measures,
synchronization measures, etc.). The appropriate choice of the
similarity measure will depend on the data. At this level the experience
(and knowledge) of the scientist collecting the data (e.g. clinician or
biologist) is valuable as it will critically guide an exchange of
information (and collaboration) with experts in TDA. Subsequently, the
emergence of simplicial complexes within the data follows a procedure
(algorithm) called \emph{Filtration} (see Fig.{~}4, Panel C), \emph{} which
consists of the following: Take as input a \emph{point cloud} (a
discrete collection of data points) and for each data point consider a
ball of radius {$\eps$~}(defined with respect to the chosen similarity
measure), centered at this data point. Subsequently, continuously
increase the radius ({$\eps$}) and monitor the intersections between the
different balls. For every intersection between n-balls, create an
n-simplex (see Fig.{~}4, Panel C). As the radius ({$\eps$}) further increases,
it generates a nested sequence of \emph{simplicial complexes} for which
the associated \emph{Euler characteristics} is computed (e.g. all types
of holes are counted in accordance to the general formula shown in see
Fig.{~}4, Panel B). Since the counting (e.g. of each type of hole) is
performed for every radius ({$\eps$}) along the generated sequence of
\emph{simplicial complexes,} a convenient way to represent this counting
process is via the so-called \emph{barcodes} (see Fig.{~}4, Panel C)~\cite{ghrist2008}.
Specifically, a \emph{barcode} (with finite length)
represents the emergence, persistence and disappearance of a given hole
along a filtration process and the set of \emph{barcodes} encodes the
entire topological signature of the data. The longest \emph{barcodes}
(those the persist the longest, therefore robust) are considered the
true topological features of the data, and the short ones represent
noise within the data. The advantage of \emph{Persistent homology} is
threefold: 1. The set of barcodes is a \emph{complete} topological
invariant, meaning that it captures all the topological information of a
filtration (i.e. a meta signature of the data across scales); 2.{~ }It
is computationally efficient; 3. It has \emph{stability} properties,
meaning that small perturbation of the data produces small perturbation
of the real topological invariant. Indeed, small holes are usually
associated with noise, while big holes are considered to be real; hence,
small pertubartions does not change the big holes much. Finally, TDA can
be used in combination with other Big Data methods, should further
processing be required. That is, the topological invariants can be
further processed for example with machine learning and statistical
methods~\cite{bergomi2019}.{~}

\begin{figure}[!b]
\centering
\includegraphics[width=0.7\textwidth]{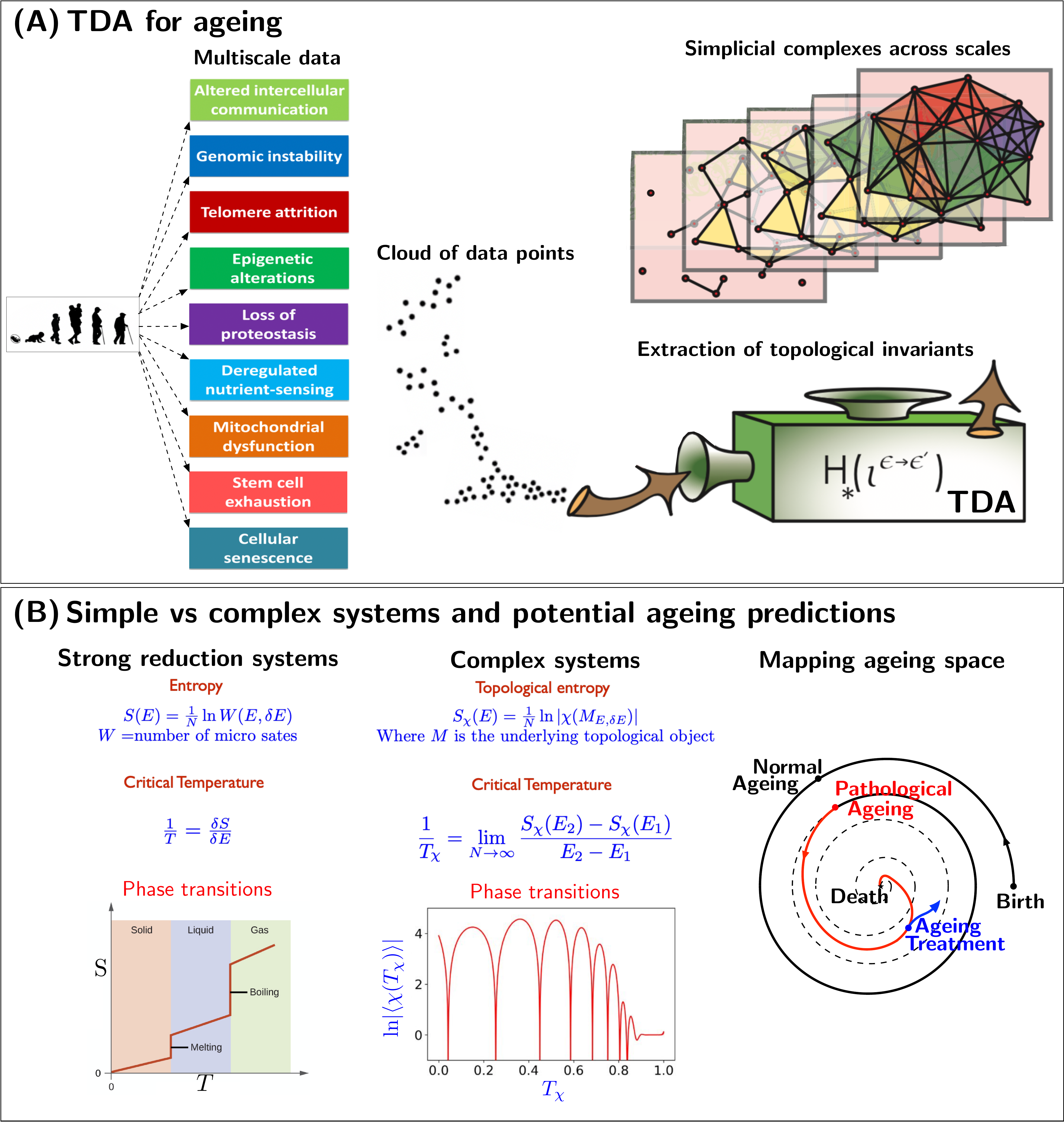}
\caption{\emph{A: proposed TDA computational pipeline to mine
topological invariances in complex high-dimensional spatial-temporal
ageing data. The analysis can be done across a heterogenous populations
and the output will reveal simplicial complexes across scales and
associated topological invariances, from which relevant clinical and
biological information can be drawn. B: A comparison between Strong
reduction system and complex systems. Recent theoretical findings show
that Topological entropies can detect multiple phase transitions in
complex data. These findings could be used to derive a map that predicts
normal ageing and associated multiscale biological mechanisms as well as
phase transitions (e.g. deviations from normal ageing). Moreover, it
could potentially suggest ageing treatments.}}
\end{figure}
%

\section*{Topological Data Analysis towards ageing data}
To the best of our knowledge, TDA has not been applied to the research
field of ageing, although it has been applied in other biological
contexts (e.g. omics data)~\cite{rabadan2019}. Therefore, this
opens a unique opportunity to add TDA to the arsenal of methods, which
in tandem would unveil unforeseen information within the complex
high-dimensional spatio-temporal ageing data. Some TDA methods (e.g.
persistent homology) are computationally efficient and scalable and thus
it is ideally suited for extracting topological invariants (e.g.
associated to evolutionary conserved mechanisms) in longitudinal ageing
data across a population. Moreover, it has the potential to identify
differences among patients, since TDA would associate. e.g., a
\emph{barcode} to each patient. Indeed, conserved mechanisms and
differences can be studied by well-defined \emph{distance measures for
barcodes}. To achieve this, we propose the TDA computational pipeline as
shown in Fig.{~}5, Panel A and a suitable similarity measure will have to
be carefully developed. The pipeline will input clinically relevant
state-of-the-art ageing data (e.g. high-throughput multi-omics) together
with other unknown biomarkers and clinical phenotypes of single patients
or of a population. Subsequently, the TDA machinery will render a
sequence of simplicial complexes (and topological invariants) across
scales. This has potential to unveil topological relationships across
multi-omics, other relevant biomarkers and clinical phenotypes.
Moreover, based on recent theoretical extensions of TDA, the topological
phase transitions (tipping points) within the data can potentially be
detected~\cite{santos2009,santos2014,santos2017,santos2019}; see Fig.{~}5, Panel B.
We will not delve into the
details of the theoretical machinery that enables the extraction these
topological phase transitions in complex systems, however, it has some
analogy with phase transitions described of strong reduction systems. As
a consequence, ageing evolution and transition (as well as disease
evolution and transition) points could be tracked via topological
invariants, which would result in an \emph{ageing map} (see Fig.{~}5, Panel
B). The envisaged map has the potential of predicting trajectories of
healthy ageing as well as mechanistically explaining multiscale
biological mechanisms, phase transitions from healthy ageing and suggest
treatments. Ultimately, the goal achievable by TDA will be to understand
the \emph{shape of ageing and age-related diseases}.

\section*{Discussion}
The complexity of ageing calls for novel sophisticated data analytical
methods that make the least amount of assumptions, yet extract global
patterns while establishing causal links between observed and latent
variables from multiscale ageing data sets. In recent years TDA has
emerged as a powerful tool that meets these aforementioned criteria and
therefore we encourage the scientific community interested in ageing to
incorporate TDA as a novel toolbox in their repertoire methods. Indeed,
the conceptual idea of TDA in analyzing data via topological means is in
stark contrast to conventional numerical summaries made by other Big
Data/AI methods, which typically make assumptions on the data and are
typically incomplete in the way causal links are established between
variables across scales. TDA extracts topological objects (simplicial
complexes) and topological invariants from data, which in itself is
novel information but also can be used as a complementary information to
that of numerical summaries made by other Big Data/AI methods. Thus, we
expect that the combination of numerical summaries (provided by other
methods such as statistics, network theory, machine learning etc.) and
topological summaries will lead to novel unprecedented insights on
ageing. Given the fast development of TDA and its numerous recent
achievements in various biological studies one can only advocate for
this method to be integrated in the toolbox that every scientist working
on ageing should have.

\hfill\break

However, a word of caution should be expressed. TDA should not be
perceived as the ultimate tool to analyse complex biological data since
the concept of ultimate tool is probably illusory. Furthermore, TDA
comes with difficulties, in particular, finding the appropriate
\emph{similarity measure} suitable to a given data set is sometimes a
daunting task and very much dependent on the knowledge of the problem
being analysed. Noteworthy, Topological approaches in theoretical
physics usually give necessary conditions, not sufficient ones~\cite{franzosi2004}.
In other words, Topology falls (most of the time)
within the class of \emph{Contextual Emergence}, since it is necessary
that two structures that are equivalent must have the same topological
properties, but the converse is not true: two structures with same
(fixed) topological property could differ in relation to another one. In
this sense, we hope that future studies of complex biological systems
and in particular ageing will reveal that the underlying mechanisms fall
under the class of \emph{Contextual Emergence.} Provided ageing falls
under Context Emergence, we envisage that TDA will help to create an
\emph{ageing map} that predicts the biological mechanisms underpinning
pathological vs healthy ageing scenarios. These phase transitions could
be unveiled in longitudinal studies but also in transversal studies
across databases, which would compare differences in the structure of
those phase transitions. In other words, TDA has the potential to
uncover the shape of ageing and age-related diseases.

\hfill\break

To conclude, TDA extracts global topological invariant patterns and it
will enable scientists and clinicians to collaborate in synergy (at the
level of systems biology) and thus significantly contribute to the
worldwide initiative on healthy ageing (WHO 2020-2030) for unraveling
the causal relationships of the ``true'' complex, dynamic, hierarchical
and emergent nature of ageing.

\bigskip 

\section*{Acknowledgement}
The works presented in the article were supported by grants from
Canadian Institutes of Health Research (CIHR) (No. 106634) and No.
PJT-162366) to TF, the Société des médecins de l'Université de
Sherbrooke and the Research Center on Aging of the CIUSSS-CHUS,
Sherbrooke and the FRQS Audace grant to TF; by {FACEPE through the
PRONEX Program (APQ-0602-1.05/14) (Brazilian agency) to FANS; }by
Ikerbasque (The Basque Foundation for Science),{~ }partially supported
by the Basque Government under the grant ``Artificial Intelligence in
BCAM number EXP. 2019/00432'', by the Basque Government through the BERC
2018-2021 program and by the Ministry of Science, Innovation and
Universities: BCAM Severo Ochoa accreditation SEV-2017-0718 and through
project RTI2018-093860B-C21 funded by (AEI/FEDER, UE) with acronym
``MathNEURO'' to SR; by Inria through Associated Team NeuroTransSF to MD
and SR.

\bigskip

\bibliographystyle{apalike}
\bibliography{refs}

\begin{thebibliography}{}

\bibitem[Baker et~al., 2018]{baker2018}
Baker, R.~E., Pe{\~n}a, J.-M., Jayamohan, J., and J{\'e}rusalem, A. (2018).
\newblock Mechanistic models versus machine learning, a fight worth fighting
  for the biological community?
\newblock {\em Biol. Lett.}, 14(5):20170660.

\bibitem[Battiston et~al., 2020]{battiston2020}
Battiston, F., Cencetti, G., Iacopini, I., Latora, V., Lucas, M., Patania, A.,
  Young, J.-G., and Petri, G. (2020).
\newblock Networks beyond pairwise interactions: structure and dynamics.
\newblock {\em Phys. Rep.}

\bibitem[Belsky et~al., 2018]{belsky2018}
Belsky, D.~W., Moffitt, T.~E., Cohen, A.~A., Corcoran, D.~L., Levine, M.~E.,
  Prinz, J.~A., Schaefer, J., Sugden, K., Williams, B., Poulton, R., et~al.
  (2018).
\newblock Eleven telomere, epigenetic clock, and biomarker-composite
  quantifications of biological aging: do they measure the same thing?
\newblock {\em Am. J. Epidemiol.}, 187(6):1220--1230.

\bibitem[Bergomi et~al., 2019]{bergomi2019}
Bergomi, M.~G., Frosini, P., Giorgi, D., and Quercioli, N. (2019).
\newblock Towards a topological--geometrical theory of group equivariant
  non-expansive operators for data analysis and machine learning.
\newblock {\em Nat. Mach. Intell.}, 1(9):423--433.

\bibitem[Bishop and Atmanspacher, 2006]{bishop2006}
Bishop, R.~C. and Atmanspacher, H. (2006).
\newblock Contextual emergence in the description of properties.
\newblock {\em Found. Phys.}, 36(12):1753--1777.

\bibitem[Bishop and Ellis, 2020]{bishop2020}
Bishop, R.~C. and Ellis, G.~F. (2020).
\newblock Contextual emergence of physical properties.
\newblock {\em Found. Phys.}, pages 1--30.

\bibitem[Blevins and Bassett, 2020]{blevins2020}
Blevins, A.~S. and Bassett, D.~S. (2020).
\newblock Reorderability of node-filtered order complexes.
\newblock {\em Phys. Rev. E}, 101(5):052311.

\bibitem[Bulterijs et~al., 2015]{bulterijs2015}
Bulterijs, S., Hull, R.~S., Bj{\"o}rk, V.~C., and Roy, A.~G. (2015).
\newblock It is time to classify biological aging as a disease.
\newblock {\em Front. Genet.}, 6:205.

\bibitem[Butterfield, 2011]{butterfield2011}
Butterfield, J. (2011).
\newblock Emergence, reduction and supervenience: A varied landscape.
\newblock {\em Found. Phys.}, 41(6):920--959.

\bibitem[Cannon, 2015]{cannon2015}
Cannon, M.~L. (2015).
\newblock What is aging?
\newblock {\em Dis. Mon.}, 61(11):454--459.

\bibitem[Carlsson, 2009]{carlsson2009}
Carlsson, G. (2009).
\newblock Topology and data.
\newblock {\em Bull. Amer. Math. Soc.}, 46(2):255--308.

\bibitem[Chmielewski, 2020]{chmielewski2020}
Chmielewski, P.~P. (2020).
\newblock Human ageing as a dynamic, emergent and malleable process: from
  disease-oriented to health-oriented approaches.
\newblock {\em Biogerontology}, 21(1):125--130.

\bibitem[Cohen et~al., 2020a]{cohen2020a}
Cohen, A., Legault, V., and F{\"u}l{\"o}p, T. (2020a).
\newblock What if there's no such thing as ``aging''?
\newblock \emph{Mech. Age. Dev.} (submitted).

\bibitem[Cohen, 2016]{cohen2016}
Cohen, A.~A. (2016).
\newblock Complex systems dynamics in aging: new evidence, continuing
  questions.
\newblock {\em Biogerontology}, 17(1):205--220.

\bibitem[Cohen et~al., 2020b]{cohen2020b}
Cohen, A.~A., Kennedy, B.~K., Anglas, U., Bronikowski, A.~M., Deelen, J.,
  Dufour, F., Ferbeyre, G., Ferrucci, L., Franceschi, C., Frasca, D., et~al.
  (2020b).
\newblock Lack of consensus on an aging biology paradigm? a global survey
  reveals an agreement to disagree, and the need for an interdisciplinary
  framework.
\newblock {\em Mech. Age. Dev.}, page 111316.

\bibitem[Cohen et~al., 2015]{cohen2015}
Cohen, A.~A., Li, Q., Milot, E., Leroux, M., Faucher, S., Morissette-Thomas,
  V., Legault, V., Fried, L.~P., and Ferrucci, L. (2015).
\newblock Statistical distance as a measure of physiological dysregulation is
  largely robust to variation in its biomarker composition.
\newblock {\em PloS one}, 10(4):e0122541.

\bibitem[da~Costa et~al., 2016]{da2016}
da~Costa, J.~P., Vitorino, R., Silva, G.~M., Vogel, C., Duarte, A.~C., and
  Rocha-Santos, T. (2016).
\newblock A synopsis on aging—theories, mechanisms and future prospects.
\newblock {\em Ageing Res. Rev.}, 29:90--112.

\bibitem[Dumont and Prakash, 2014]{dumont2014}
Dumont, S. and Prakash, M. (2014).
\newblock Emergent mechanics of biological structures.
\newblock {\em Mol. Biol. Cell}, 25(22):3461--3465.

\bibitem[Ehrenfest and Ehrenfest, 1959]{ehrenfest1959}
Ehrenfest, P. and Ehrenfest, T. (1959).
\newblock {\em The Conceptual Foundations of the Statistical Approach in
  Mechanics}.
\newblock Cornell Univ. Press.

\bibitem[Franceschi et~al., 2017]{franceschi2017}
Franceschi, C., Salvioli, S., Garagnani, P., de~Eguileor, M., Monti, D., and
  Capri, M. (2017).
\newblock Immunobiography and the heterogeneity of immune responses in the
  elderly: a focus on inflammaging and trained immunity.
\newblock {\em Front. Immunol.}, 8:982.

\bibitem[Franzosi and Pettini, 2004]{franzosi2004}
Franzosi, R. and Pettini, M. (2004).
\newblock Theorem on the origin of phase transitions.
\newblock {\em Phys. Rev. Lett.}, 92(6):060601.

\bibitem[F{\"u}l{\"o}p et~al., 2016]{fulop2016}
F{\"u}l{\"o}p, T., Dupuis, G., Baehl, S., Le~Page, A., Bourgade, K., Frost, E.,
  Witkowski, J., Pawelec, G., Larbi, A., and Cunnane, S. (2016).
\newblock From inflamm-aging to immune-paralysis: a slippery slope during aging
  for immune-adaptation.
\newblock {\em Biogerontology}, 17(1):147--157.

\bibitem[F{\"u}l{\"o}p et~al., 2018]{fulop2018}
F{\"u}l{\"o}p, T., Larbi, A., Dupuis, G., Le~Page, A., Frost, E.~H., Cohen,
  A.~A., Witkowski, J.~M., and Franceschi, C. (2018).
\newblock Immunosenescence and inflamm-aging as two sides of the same coin:
  friends or foes?
\newblock {\em Front. Immunol.}, 8:1960.

\bibitem[F{\"u}l{\"o}p et~al., 2019]{fulop2019}
F{\"u}l{\"o}p, T., Larbi, A., Khalil, A., Cohen, A.~A., and Witkowski, J.~M.
  (2019).
\newblock Are we ill because we age?
\newblock {\em Front. Physiol.}, 10.

\bibitem[Geniesse et~al., 2019]{geniesse2019}
Geniesse, C., Sporns, O., Petri, G., and Saggar, M. (2019).
\newblock Generating dynamical neuroimaging spatiotemporal representations
  (dyneusr) using topological data analysis.
\newblock {\em Network Neuroscience}, 3(3):763--778.

\bibitem[Ghrist, 2008]{ghrist2008}
Ghrist, R. (2008).
\newblock Barcodes: the persistent topology of data.
\newblock {\em Bull. Amer. Math. Soc.}, 45(1):61--75.

\bibitem[Gottlieb, 2003]{gottlieb2003}
Gottlieb, A.~D. (2003).
\newblock Propagation of chaos in classical and quantum kinetics.
\newblock In {\em Stochastic Analysis and Mathematical Physics II}, pages
  135--146. Springer.

\bibitem[Haag et~al., 1967]{haag1967}
Haag, R., Hugenholtz, N.~M., and Winnink, M. (1967).
\newblock On the equilibrium states in quantum statistical mechanics.
\newblock {\em Commun. Math. Phys.}, 5(3):215--236.

\bibitem[Hatcher, 2002]{hatcher2002}
Hatcher, A. (2002).
\newblock {\em Algebraic Topology}.
\newblock Cambridge University Press.

\bibitem[Hinks et~al., 2015]{hinks2015}
Hinks, T., Zhou, X., Staples, K., Dimitrov, B., Manta, A., Petrossian, T., Lum,
  P., Smith, C., Ward, J., Howarth, P., et~al. (2015).
\newblock Multidimensional endotypes of asthma: topological data analysis of
  cross-sectional clinical, pathological, and immunological data.
\newblock {\em The Lancet}, 385:S42.

\bibitem[Hinks et~al., 2016]{hinks2016}
Hinks, T.~S., Brown, T., Lau, L.~C., Rupani, H., Barber, C., Elliott, S., Ward,
  J.~A., Ono, J., Ohta, S., Izuhara, K., et~al. (2016).
\newblock Multidimensional endotyping in patients with severe asthma reveals
  inflammatory heterogeneity in matrix metalloproteinases and chitinase 3--like
  protein 1.
\newblock {\em J. Allergy Clin. Immunol.}, 138(1):61--75.

\bibitem[Horvath, 2013]{horvath2013}
Horvath, S. (2013).
\newblock Dna methylation age of human tissues and cell types.
\newblock {\em Genome Biol.}, 14(10):3156.

\bibitem[Horvath and Raj, 2018]{horvath2018}
Horvath, S. and Raj, K. (2018).
\newblock Dna methylation-based biomarkers and the epigenetic clock theory of
  ageing.
\newblock {\em Nat. Rev. Genet.}, 19(6):371.

\bibitem[Kennedy et~al., 2014]{kennedy2014}
Kennedy, B.~K., Berger, S.~L., Brunet, A., Campisi, J., Cuervo, A.~M., Epel,
  E.~S., Franceschi, C., Lithgow, G.~J., Morimoto, R.~I., Pessin, J.~E., et~al.
  (2014).
\newblock Geroscience: linking aging to chronic disease.
\newblock {\em Cell}, 159(4):709--713.

\bibitem[Kivel{\"a} et~al., 2014]{kivela2014}
Kivel{\"a}, M., Arenas, A., Barthelemy, M., Gleeson, J.~P., Moreno, Y., and
  Porter, M.~A. (2014).
\newblock Multilayer networks.
\newblock {\em J. Compl. Netw.}, 2(3):203--271.

\bibitem[Kubo, 1957]{kubo1957}
Kubo, R. (1957).
\newblock General theory and simple applications to magnetic and conduction
  problems.
\newblock {\em J. Phys. Soc. Jpn}, 12:570.

\bibitem[Lambiotte et~al., 2019]{lambiotte2019}
Lambiotte, R., Rosvall, M., and Scholtes, I. (2019).
\newblock From networks to optimal higher-order models of complex systems.
\newblock {\em Nat. Phys.}, 15(4):313--320.

\bibitem[LeCun et~al., 2015]{lecun2015}
LeCun, Y., Bengio, Y., and Hinton, G. (2015).
\newblock Deep learning.
\newblock {\em Nature}, 521(7553):436--444.

\bibitem[Li et~al., 2015]{li2015}
Li, L., Cheng, W.-Y., Glicksberg, B.~S., Gottesman, O., Tamler, R., Chen, R.,
  Bottinger, E.~P., and Dudley, J.~T. (2015).
\newblock Identification of type 2 diabetes subgroups through topological
  analysis of patient similarity.
\newblock {\em Sci. Transl. Med.}, 7(311):311ra174--311ra174.

\bibitem[Li et~al., 2020]{li2020}
Li, X., Ploner, A., Wang, Y., Magnusson, P.~K., Reynolds, C., Finkel, D.,
  Pedersen, N.~L., Jylh{\"a}v{\"a}, J., and H{\"a}gg, S. (2020).
\newblock Longitudinal trajectories, correlations and mortality associations of
  nine biological ages across 20-years follow-up.
\newblock {\em Elife}, 9:e51507.

\bibitem[Lipsky and King, 2015]{lipsky2015}
Lipsky, M.~S. and King, M. (2015).
\newblock Biological theories of aging.
\newblock {\em Dis. Mon.}, 61(11):460.

\bibitem[L{\'o}pez-Ot{\'\i}n et~al., 2013]{lopez2013}
L{\'o}pez-Ot{\'\i}n, C., Blasco, M.~A., Partridge, L., Serrano, M., and
  Kroemer, G. (2013).
\newblock The hallmarks of aging.
\newblock {\em Cell}, 153(6):1194--1217.

\bibitem[Martin and Schwinger, 1959]{martin1959}
Martin, P.~C. and Schwinger, J. (1959).
\newblock Theory of many-particle systems. i.
\newblock {\em Phys. Rev.}, 115(6):1342.

\bibitem[Milot et~al., 2014]{milot2014}
Milot, E., Morissette-Thomas, V., Li, Q., Fried, L.~P., Ferrucci, L., and
  Cohen, A.~A. (2014).
\newblock Trajectories of physiological dysregulation predicts mortality and
  health outcomes in a consistent manner across three populations.
\newblock {\em Mech. Ageing Dev.}, 141:56--63.

\bibitem[Mitnitski et~al., 2017]{mitniski2017}
Mitnitski, A., Rutenberg, A., Farrell, S., and Rockwood, K. (2017).
\newblock Aging, frailty and complex networks.
\newblock {\em Biogerontology}, 18(4):433--446.

\bibitem[Moskalev, 2020]{moskalev2020}
Moskalev, A. (2020).
\newblock The challenges of estimating biological age.
\newblock {\em eLife}, 9.

\bibitem[Petri et~al., 2013]{petri2013}
Petri, G., Scolamiero, M., Donato, I., and Vaccarino, F. (2013).
\newblock Topological strata of weighted complex networks.
\newblock {\em PloS one}, 8(6):e66506.

\bibitem[Rabad{\'a}n and Blumberg, 2019]{rabadan2019}
Rabad{\'a}n, R. and Blumberg, A.~J. (2019).
\newblock {\em Topological Data Analysis for Genomics and Evolution: Topology
  in Biology}.
\newblock Cambridge University Press.

\bibitem[Rose et~al., 2012]{rose2012}
Rose, M., Flatt, T., Graves, J., Greer, L., Martinez, D., Matos, M., Mueller,
  L., Shmookler~Reis, R., and Shahrestani, P. (2012).
\newblock What is aging?
\newblock {\em Front. Genet.}, 3:134.

\bibitem[Sagiroglu and Sinanc, 2013]{sagiroglu2013}
Sagiroglu, S. and Sinanc, D. (2013).
\newblock Big data: A review.
\newblock In {\em 2013 international conference on collaboration technologies
  and systems (CTS)}, pages 42--47. IEEE.

\bibitem[Salnikov et~al., 2018]{salnikov2018}
Salnikov, V., Cassese, D., and Lambiotte, R. (2018).
\newblock Simplicial complexes and complex systems.
\newblock {\em Eur. J. Phys.}, 40(1):014001.

\bibitem[Santos and Coutinho-Filho, 2009]{santos2009}
Santos, F. and Coutinho-Filho, M. (2009).
\newblock Topology, symmetry, phase transitions, and noncollinear spin
  structures.
\newblock {\em Phys. Rev. E}, 80(3):031123.

\bibitem[Santos et~al., 2017]{santos2017}
Santos, F., da~Silva, L., and Coutinho-Filho, M. (2017).
\newblock Topological approach to microcanonical thermodynamics and phase
  transition of interacting classical spins.
\newblock {\em J. Stat. Mech.: Theory Exp.}, 2017(1):013202.

\bibitem[Santos et~al., 2014]{santos2014}
Santos, F., Rehn, J., and Coutinho-Filho, M. (2014).
\newblock Topological and geometrical aspects of phase transitions.
\newblock In {\em Perspectives and Challenges in Statistical Physics and
  Complex Systems for the Next Decade}, pages 153--180. World Scientific.

\bibitem[Santos et~al., 2019]{santos2019}
Santos, F.~A., Raposo, E.~P., Coutinho-Filho, M.~D., Copelli, M., Stam, C.~J.,
  and Douw, L. (2019).
\newblock Topological phase transitions in functional brain networks.
\newblock {\em Phys. Rev. E}, 100(3):032414.

\bibitem[Sasaki et~al., 2020]{sasaki2020}
Sasaki, K., Bruder, D., and Hernandez-Vargas, E.~A. (2020).
\newblock Topological data analysis to model the shape of immune responses
  during co-infections.
\newblock {\em Commun. Nonlinear Sci.}, 85:105228.

\bibitem[Singh et~al., 2007]{singh2007}
Singh, G., M{\'e}moli, F., and Carlsson, G.~E. (2007).
\newblock Topological methods for the analysis of high dimensional data sets
  and 3d object recognition.
\newblock {\em SPBG}, 91:100.

\bibitem[Torres et~al., 2016]{torres2016}
Torres, B.~Y., Oliveira, J. H.~M., Thomas~Tate, A., Rath, P., Cumnock, K., and
  Schneider, D.~S. (2016).
\newblock Tracking resilience to infections by mapping disease space.
\newblock {\em PLoS Biol}, 14(4):e1002436.

\bibitem[Villani, 2002]{villani2002}
Villani, C. (2002).
\newblock A review of mathematical topics in collisional kinetic theory.
\newblock {\em Handbook of Mathematical Fluid Dynamics}, 1(71-305):3--8.

\bibitem[Wasserman, 2018]{wasserman2018}
Wasserman, L. (2018).
\newblock Topological data analysis.
\newblock {\em Ann. Rev. Stat. Appl.}, 5:501--532.

\bibitem[Whitwell et~al., 2020]{whitwell2020}
Whitwell, H.~J., Bacalini, M.~G., Blyuss, O., Chen, S., Garagnani, P.,
  Gordleeva, S.~Y., Jalan, S., Ivanchenko, M., Kanakov, O., Kustikova, V.,
  et~al. (2020).
\newblock The human body as a super network: Digital methods to analyze the
  propagation of aging.
\newblock {\em Front. Aging Neurosci.}, 12:136.

\bibitem[Zhavoronkov et~al., 2019]{zhavoronkov2019}
Zhavoronkov, A., Mamoshina, P., Vanhaelen, Q., Scheibye-Knudsen, M., Moskalev,
  A., and Aliper, A. (2019).
\newblock Artificial intelligence for aging and longevity research: Recent
  advances and perspectives.
\newblock {\em Ageing Res. Rev.}, 49:49--66.

\bibitem[Zomorodian, 2005]{zomorodian2005}
Zomorodian, A.~J. (2005).
\newblock {\em Topology for computing}.
\newblock Cambridge University Press.

\end{thebibliography}
\end{document}